\documentclass[journal]{IEEEtran}

\ifCLASSINFOpdf
\else
   \usepackage[dvips]{graphicx}
\fi
\usepackage{url}
\usepackage{cite}
\usepackage{amsmath,amssymb,amsfonts}
\usepackage{algorithmic}
\usepackage{graphicx}
\usepackage{textcomp}
\usepackage{xcolor}
\usepackage{rotating}
\usepackage{cite}
\usepackage{booktabs}
\usepackage{algorithm}
\usepackage{multirow}
\usepackage{mathrsfs}
\usepackage{graphicx}
\usepackage{hyperref}
\usepackage{subcaption}
\usepackage{amsmath}

\begin{document}

\title{Bit-depth color recovery via off-the-shelf super-resolution models}

\author{Xuanshuo Fu, Danna Xue, and Javier Vazquez-Corral


\thanks{This paper was supported by Grant PID2021-128178OB-I00 funded by MCIN/AEI/10.13039/501100011033, ERDF ``A way of making Europe", the Departament de Recerca i Universitats from Generalitat de Catalunya with reference 2021SGR01499. X. Fu is supported by the predoctoral program AGAUR-FI ajuts (2024 FI-3 00065) Joan Oró, which is backed by the Secretariat of Universities and Research of the Department of Research and Universities of the Generalitat of Catalonia, as well as the European Social Plus Fund.}
\thanks{X. Fu, D. Xue and J. Vazquez-Corral are with Computer Vision Center \& Universitat Autònoma de Barcelona, Barcelona, Spain (e-mail: \{xuanshuo, dxue, jvazquez\}@cvc.uab.cat).}}

\markboth{Journal of \LaTeX\ Class Files, Vol. 14, No. 8, August 2015}
{Shell \MakeLowercase{\textit{et al.}}: Bare Demo of IEEEtran.cls for IEEE Journals}
\maketitle

\begin{abstract}
Advancements in imaging technology have enabled hardware to support 10 to 16 bits per channel, facilitating precise manipulation in applications like image editing and video processing. While deep neural networks promise to recover high bit-depth representations, existing methods often rely on scale-invariant image information, limiting performance in certain scenarios. In this paper, we introduce a novel approach that integrates a super-resolution architecture to extract detailed a priori information from images. By leveraging interpolated data generated during the super-resolution process, our method achieves pixel-level recovery of fine-grained color details. Additionally, we demonstrate that spatial features learned through the super-resolution process significantly contribute to the recovery of detailed color depth information. Experiments on benchmark datasets demonstrate that our approach outperforms state-of-the-art methods, highlighting the potential of super-resolution for high-fidelity color restoration.
\end{abstract}


\IEEEpeerreviewmaketitle

\section{Introduction}





The continuous advancements in imaging technology and hardware have driven the widespread adoption of devices supporting 10- to 16-bit color representations~\cite{apple_support_sp770, samsung_s10_specs}. These advancements have significantly raised the standards for image and video processing, enabling finer color adjustments and enhanced detail preservation. However, a substantial portion of digital content still relies on 8-bit color representation~\cite{karaimer2016software}, which restricts advanced editing capabilities and hampers the preservation of fine details in imaging workflows. As the demand for higher-quality imaging grows, bit-depth recovery, which transforms lower bit-depth images into higher bit-depth counterparts, has emerged as a crucial area of research. This process not only enhances visual quality but also ensures better compatibility with modern imaging devices.

Traditional bit-depth recovery methods, such as gain factor multiplication~\cite{ulichney1998pixel}, bit replication~\cite{ulichney1998pixel}, contour reconstruction~\cite{cheng2009bit}, and optimization-based techniques~\cite{mittal2012bit, wan20122d, wan2016image, liu2018ipad}, are computationally efficient but often fail to retain intricate details, resulting in banding artifacts and texture loss. In contrast, recent approaches leverage deep neural networks, such as UNet-based~\cite{byun2019bitnet} architectures, and other richer feature representations~\cite{su2019photo, liu2019calf, punnappurath2021little, zhao2019deep, hou2017image} to achieve fine-grained color recovery. BitMore~\cite{punnappurath2021little} introduces the idea of performing bit-depth recovery in the binary space by predicting the higher bits step-by-step. This approach allows for flexible depth restoration by employing different submodels during inference. However, many methods, including BitMore~\cite{punnappurath2021little}, primarily focus on single-scale features and fail to account for the multi-scale nature of image information. This limitation reduces their effectiveness in capturing the spatial details necessary for accurately restoring textures, edges, and fine patterns.

The image super-resolution (SR) task reconstructs image details to increase spatial resolution~\cite{SR_survey}. The color of a pixel is strongly correlated with neighboring pixel values, particularly in smooth image regions. SR approaches leverage this contextual information to recover high-resolution spatial details, while bit-depth recovery similarly relies on contextual cues to achieve more precise color depth representation. Building on these similarities, we propose a novel method that integrates SR techniques to enhance bit-depth recovery. Specifically, our approach incorporates a pre-trained SR encoder as a preprocessing module to extract fine-grained spatial features, facilitating detailed bit-depth reconstruction. By using an off-the-shelf encoder with pre-trained weights, we eliminate the need for additional training. Unlike existing methods, our framework explicitly employs multi-scale feature extraction to enhance texture and edge recovery while effectively leveraging contextual information. This design ensures higher accuracy and improved image fidelity in bit-depth recovery. Our main contribution includes:

\begin{itemize} 
\item We propose to use pre-trained super-resolution encoders to extract spatial priors, enabling precise recovery of fine-grained details during bit-depth recovery.
\item Our method captures multi-resolution contextual information, addressing limitations of fixed-scale approaches.
\item  Experiments on four benchmarks demonstrate that our method outperforms traditional and deep learning-based techniques in PSNR, SSIM, and visual quality.
\end{itemize}


\section{Methodology}

\begin{figure*}[t!]
  \centering
  \includegraphics[width=0.95\linewidth]{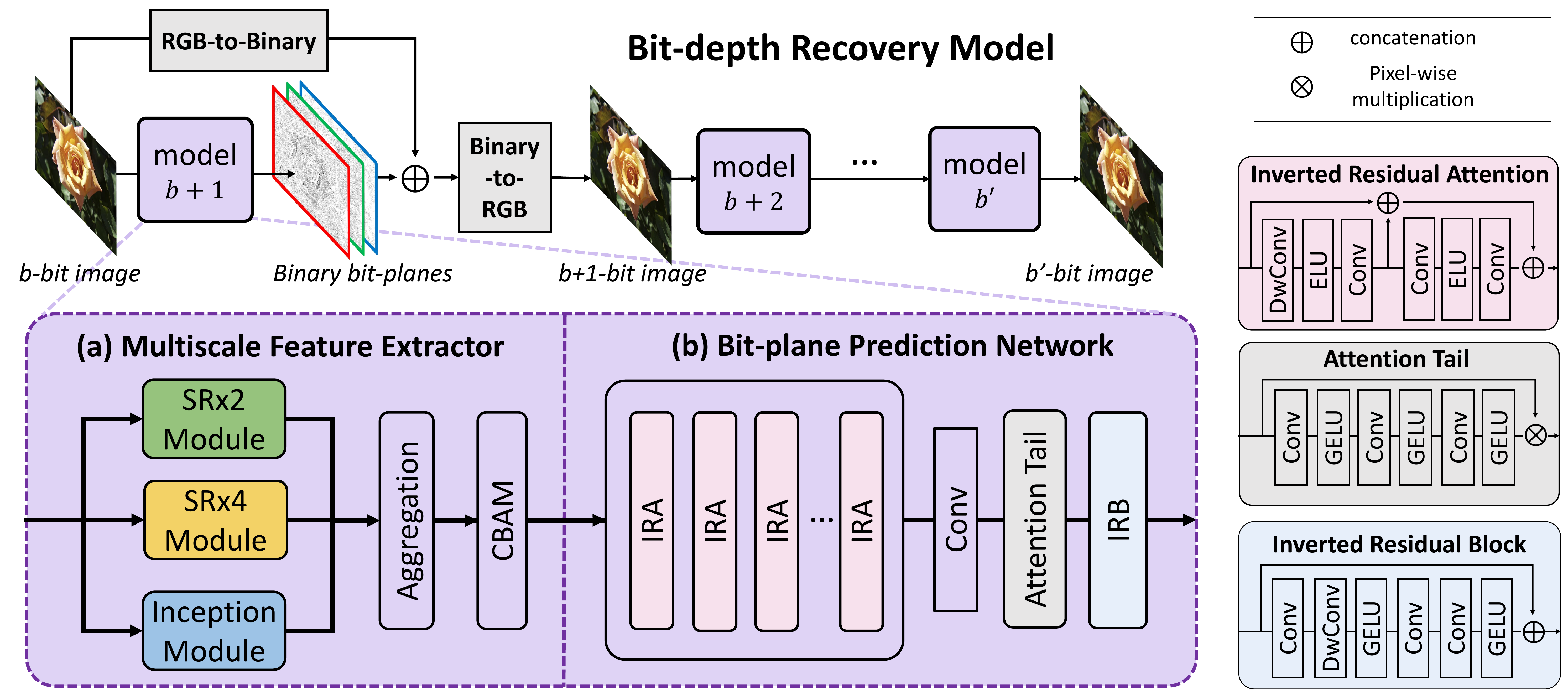}
  \caption{The framework of our approach. The bit-depth recovery model comprises several submodels with the same architecture but different weights, and recovers the color depth bit-by-bit with each submodel. Given a b-bit image, the submodel predicts three binary bit-planes for R, G, and B channels, respectively. The predicted bit-plane is concatenated with the binary bit-planes of the input image and then mapped back to the b+1-bit values. By processing the image step by step, we can obtain the final target image. Each submodel includes two parts: (a) the multi-scale feature encoder and (b) the bit-plane prediction network. The multi-scale feature encoder consists of several super-resolution encoders pretrained for different scale SR tasks, an inception module, and a feature aggregation module that fuses the features. The multi-scale features are then processed by the bit-plane prediction network including different blocks to predict the output binary bit-planes.}
  \label{fig:net}
\end{figure*}

In this section, we introduce a new architecture for bit-depth recovery, as illustrated in Fig.~\ref{fig:net}. The architecture comprises two key components: (a) SR-based multi-scale feature extractor (Sec.~\ref{sec:extractor}) and (b) Attention-augmented bit-plane recovery network (Sec.~\ref{sec:baseline}). The former component utilizes a pre-trained super-resolution feature extractor to extract and fuse multi-scale image features, which capture more detailed image information and higher resolution. The latter is a lightweight network designed to progressively recover information from each bit-plane, enabling the restoration of fine details and complex textures.

\subsection{Overview}
\label{sec:overview}

Given a low-bit RGB image $I_{L}$ with bit depth $b_{L}$, where each pixel value of $I_{L}$ is within the range of $[0, 2^{b_{L}} - 1]$. The target of bit-depth recovery task is to restore a higher-bit RGB image $I_{H}$ with each pixel value $p' \in [0, 2^{b_{H}} - 1],~b_{H}>b_{L}$. 

As shown in Fig.~\ref{fig:net}, our bit-depth recovery model $\Phi$ recovers the color depth in a bit-by-bit manner~\cite{punnappurath2021little}. The model includes several submodels with the same architecture but different weights, $\Phi=\{ \phi_{b+1}, \phi_{b+2}, ..., \phi_{b'} \}$. Given a b-bit image $I_{b}$, the submodel $\phi_{b+1}$ predicts the ${b+1}^{th}$ binary bit-plane $P_{b+1}$ by
\begin{equation}
    P_{b+1} = \phi_{b+1}(I_{b}).
\end{equation}
By combining this predicted bit-plane with the bit-planes of the input image, we can obtain the b+1-bit image $I_{b+1}$.

We use the Binary Cross Entropy (BCE) loss for each submodel training, which is given by:

\begin{equation}
    \mathcal{L} = \sum_{n=1}^{N} y_n \cdot log\sigma(x_n) + (1 - y_n) \cdot log(1 - \sigma(x_n)),
\end{equation}

\noindent where $x_n$ is the binary output of each submodel, $y_n$ is the groundtruth of the binary bit-plane and $N$ is the number of samples. $\sigma$ is the sigmoid activation function.

\begin{table*}[t!]
    \caption{Results on Sintel dataset~\cite{Sintel}. Results reported are the mean for all the images.}
    \centering
    \small
    \resizebox{\textwidth}{!}{
    \begin{tabular}{ccccccccccccc}
    \toprule
    \multirow{2}{*}{\textbf{Method}} & \multicolumn{2}{c}{\textbf{4-16 bit}} & \multicolumn{2}{c}{\textbf{4-12 bit}} & \multicolumn{2}{c}{\textbf{4-8 bit}} & \multicolumn{2}{c}{\textbf{6-16 bit}} & \multicolumn{2}{c}{\textbf{6-12 bit}} & \multicolumn{2}{c}{\textbf{8-16 bit}}\\
     & PSNR & SSIM & PSNR & SSIM & PSNR & SSIM & PSNR & SSIM & PSNR & SSIM & PSNR & SSIM\\
    \midrule
    \textbf{BE-CALF}~\cite{liu2019calf} & 39.9829 & 0.9752 & 39.9840 & 0.9752 & 39.9072 & 0.9737 & \color{black}{51.1430} & \color{black}{0.9940} & \color{black}{51.1454} & \color{black}{0.9940} & 59.5117 & 0.9993 \\
    \textbf{BitNet}~\cite{byun2019bitnet} & 39.4893 & 0.9719 & 39.4931 & 0.9719 & 39.3369 & 0.9701 & 49.6795 & 0.9954 & 49.7192 & 0.9954 & 57.5487 & 0.9989 \\
    \midrule
    \textbf{BitMore D4}~\cite{punnappurath2021little} & 40.9274 & 0.9786 & 40.9286 & 0.9786 & 40.6143 & 0.9773 & 52.7599 & 0.9976 & 52.7491 & 0.9976 & 63.0731 & 0.9997 \\
    \textbf{Ours-4} & \textbf{41.3400} & \textbf{0.9813} & \textbf{41.3416} & \textbf{0.9813} & \textbf{41.009} & \textbf{0.9793} & \textbf{53.2326} & \textbf{0.9980} & \textbf{53.2101} & \textbf{0.9980} & \textbf{63.3795} & \textbf{0.9998} \\
    \midrule
    \textbf{BitMore D16}~\cite{punnappurath2021little} & 41.5070 & 0.9810 & 41.5080 & 0.9810 & 41.1909 & 0.9794 & 53.4825 & 0.9979 & 53.4731 & 0.9980 & 63.5146 & \textbf{0.9998} \\
    \textbf{Ours-16} & \textbf{42.0072} & \textbf{0.9833} & \textbf{42.0009} & \textbf{0.9833} & \textbf{41.5245} & \textbf{0.9812} & \textbf{53.7273} & \textbf{0.9982} & \textbf{53.7092} & \textbf{0.9982} & \textbf{63.5633} & \textbf{0.9998} \\
    \bottomrule
    \end{tabular}
    }
    \label{tab:se}
\end{table*}

\subsection{SR-based Multi-scale Feature Extractor}
\label{sec:extractor}

The multiscale feature extractor is a key component of the proposed architecture, leveraging multi-resolution information for high-fidelity color restoration. The feature extractor consists of three paths: two SR feature extraction modules operating at different scales and one inception-based module that works on the input image scale. The outputs of these paths are aggregated and passed through the Convolutional Block Attention Module (CBAM). This module assigns importance scores to both spatial and channel-wise features, prioritizing the most relevant information for downstream processing.

For the SR feature extraction modules, we employ two super-resolution feature encoders trained with datasets of different scales, specifically $\times2$ and $\times4$, respectively. These SR modules are taken from widely used super-resolution architectures, \textit{e.g.}, the EDSR~\cite{lim2017enhanced} model. We directly use the pre-trained weights trained on SR datasets, which contain images different from those in our bit-depth datasets, to retain their originally learned feature representations. This approach ensures that the SR modules provide robust and domain-agnostic priors without introducing the risk of overfitting or adding additional training complexity. By excluding the upsampling operation at the end of the original SR network, the SR modules in our feature extractor can obtain the fine-grained priors learned from the SR task while avoiding extra computational costs due to spatially high-resolution features.

The SR model facilitates the recovery of missing color details by refining spatial relationships and reconstructing gradients across channels. This process enhances the accuracy of color mapping by ensuring continuity and consistency across pixels, which is particularly beneficial for high-bit-depth data where subtle variations in tone and hue are critical. Additionally, the three modules extract features at different scales, ranging from coarse global patterns to fine-grained local details. This hierarchical representation enhances both the lower bits, focusing on low-frequency structures, and the higher bits, focusing on textures and details. Moreover, the frozen SR modules provide domain-agnostic priors that generalize well to color restoration tasks. With these priors, our model effectively extracts spatial features that contribute to high-fidelity restoration of color depth information.

The inception block utilizes multiscale convolutional filters to capture features across varying levels of granularity. This module complements the SR modules by providing broader contextual information. This multiscale architecture enables the network to balance local detail extraction with global context understanding. Note that, unlike the SR modules, this module is trained from scratch.

\subsection{Attention-augmented Bit-plane Prediction Network}
\label{sec:baseline}

After obtaining the aggregated features, our bit-plane prediction network processes the refined representations through a series of Inverted Residual Attention (IRA) modules~\cite{conde2023perceptual}, which incorporate lightweight convolutions and attention mechanisms to enhance feature quality. The IRA blocks utilize residual connections to retain critical information while dynamically adjusting feature importance.

The Attention Tail Module further refines features by emphasizing relevant spatial regions or feature channels, thereby improving the model's capacity to restore fine details. Finally, the output passes through an Inverted Residual Block (IRB)~\cite{conde2023perceptual}, which balances color distribution and restores tonal information, ensuring that the final output maintains both local detail and global consistency.


\section{Experiments}

\subsection{Datasets}
We use five different datasets for model training and testing. Among them, the Sintel dataset~\cite{Sintel}, and TESTIMAGES~\cite{asuni2014testimages} provide 16-bit images, allowing us to present six bit-depth recovery settings, ranging from 4-to-8 bit to 8-to-16 bit. Since the Kodak~\cite{Kodak} and ESPL v2~\cite{kundu2015full} datasets only contain 8-bit images, we present two settings: 3-to-8 bit and 4-to-8 bit. In the experiments, low-bit images are generated from the original higher-bit images by quantization to serve as model inputs. We randomly select 1,000 images from each of the MIT-Adobe 5K and Sintel datasets to build a joint training set. These selected images are consistent with the training set used by BitMore~\cite{punnappurath2021little}. For testing, we adhere to the evaluation protocol established by BitMore, ensuring comparability with previous methods. Our model is evaluated on the test sets of the Sintel, TESTIMAGES, Kodak, and ESPL v2 datasets.

\begin{table*}[t!]
    \caption{Results on TESTIMAGES 1200 dataset~\cite{asuni2014testimages}}
    \centering
    \small
    \resizebox{\textwidth}{!}{
    \begin{tabular}{ccccccccccccc}
    \toprule
    \multirow{2}{*}{\textbf{Method}}  & \multicolumn{2}{c}{\textbf{4-16 bit}} & \multicolumn{2}{c}{\textbf{4-12 bit}} & \multicolumn{2}{c}{\textbf{4-8 bit}} & \multicolumn{2}{c}{\textbf{6-16 bit}} & \multicolumn{2}{c}{\textbf{6-12 bit}} & \multicolumn{2}{c}{\textbf{8-16 bit}}\\
     & PSNR & SSIM & PSNR & SSIM & PSNR & SSIM & PSNR & SSIM & PSNR & SSIM & PSNR & SSIM\\
    \midrule
    
    \textbf{BE-CALF}~\cite{liu2019calf} & 38.5099 & 0.9649 & 38.5095 & 0.9648 & 38.4572 & 0.9632 & \color{black}{49.8488} & \color{black}{0.9945} & \color{black}{49.8521} & \color{black}{0.9945} & 58.1167 & 0.9992 \\
    \textbf{BitNet}~\cite{byun2019bitnet} & 38.8073 & 0.9589 & 38.8158 & 0.9589 & 38.7515 & 0.9571 & 49.4834 & 0.9944 & 49.5259 & 0.9944 & 53.6031 & 0.9970 \\
    \midrule
    \textbf{BitMore D4}~\cite{punnappurath2021little} & 39.6503 & 0.9700 & 39.6619 & 0.9700 & 39.6822 & 0.9691 & 51.5413 & 0.9964 & 51.5490 & 0.9964 & 61.3626 & \textbf{0.9996} \\
    \textbf{Ours-4} & \textbf{39.8284} & \textbf{0.9717} & \textbf{39.8379} & \textbf{0.9717} & \textbf{39.7858} & \textbf{0.9697} & \textbf{51.8430} & \textbf{0.9967} & \textbf{51.8349} & \textbf{0.9967} & \textbf{61.5903} & \textbf{0.9996}\\
    \midrule
    \textbf{BitMore D16}~\cite{punnappurath2021little} & 40.4099 & 0.9735 & 40.4216 & 0.9735 & 40.3906 & 0.9725 & 52.1204 & 0.9967 & 52.1220 & 0.9967 & \textbf{61.6839} & \textbf{0.9996} \\
    \textbf{Ours-16} & \textbf{40.7007} & \textbf{0.9749} & \textbf{40.7046} & \textbf{0.9749} & \textbf{40.5228} & \textbf{0.9734} & \textbf{52.1957} & \textbf{0.9969} & \textbf{52.1864} & \textbf{0.9969} & 61.6612 & \textbf{0.9996}\\
    \bottomrule
    \end{tabular}
    }
    \label{tab:te}
\end{table*}

\subsection{Implementation Details}

We train each submodel for a total of 200 epochs, using the Stochastic Gradient Descent (SGD) optimizer for the first 50 epochs with a learning rate of 0.001, a momentum of 0.9, and a decay rate of 0.0001. For the remaining 150 epochs, we switch to the Adam optimizer, also with a learning rate of 0.001 and a decay rate of 0.0001. All training and testing are conducted on an NVIDIA 3090 GPU.

Our experiments include models employing the feature extraction components of EDSR~\cite{lim2017enhanced} and RCAN~\cite{zhang2018image} as the SR modules, specifically all the blocks preceding the final upsampling layer. The EDSR and RCAN models are pretrained on DIV2K datasets~\cite{timofte2017ntire}.

\subsection{Quantitative Results}

We evaluate two version of our bit-depth recovery methods, each using a submodel containing 4 IRA blocks (Ours-4, 6.35 M parameters) and 16 IRA blocks (Ours-16, 8.15 M parameters). The small and large model are tailored to meet different performance requirements in the experiments. We present comparisons of PSNR and SSIM between the proposed methods and existing state-of-the-art bit-depth recovery approaches, including BE-CALF~\cite{liu2019calf}, BitNet~\cite{byun2019bitnet}, and BitMore~\cite{punnappurath2021little}, across four test datasets. The results demonstrate that our methods outperform previous approaches, such as BE-CALF and BitNet, showcasing their ability to recover fine-grained details from low-bit-depth images.

On the Sintel dataset (Table~\ref{tab:se}), Ours-16 achieves the highest PSNR of 42.0072 dB and an SSIM of 0.9833 in the 4-to-16 bit conversion, significantly surpassing BitMore D16. Additionally, in the 8-to-16 bit conversion, both Ours-4 and Ours-16 record an SSIM of 0.9998, demonstrating strong structural preservation and detail recovery. Similarly, Table~\ref{tab:te} highlights the superior performance of Ours-4 and Ours-16 on the TESTIMAGES 1200 dataset. Notably, Ours-16 achieves a PSNR of 40.7007 dB and an SSIM of 0.9749 in the 4-to-16 bit conversion, outperforming BitMore D16 and showcasing the robustness of our approach across multiple datasets.

Table~\ref{tab:ko} summarizes results on the Kodak dataset, where Ours-4 achieves a PSNR of 33.7392 dB and an SSIM of 0.9315 in the 3-to-8 bit conversion, surpassing BitMore D4. Ours-16 also leads in the 4-to-8 bit category with a PSNR of 39.6788 dB and an SSIM of 0.9725. Table~\ref{tab:es} presents results on the ESPL v2 dataset, showing competitive performance for Ours-4 and Ours-16 compared to BitMore D4 and D16.

\begin{table}[t!]
    \caption{Results on Kodak dataset~\cite{Kodak}. NR denotes the score is `not reported' in the original paper.}
    \centering
    \small
    \scalebox{0.9}{
    \begin{tabular}{ccccc}
    \toprule
    \multirow{2}{*}{\textbf{Method}} & \multicolumn{2}{c}{\textbf{3-8 bit}} & \multicolumn{2}{c}{\textbf{4-8 bit}}\\
    & PSNR & SSIM & PSNR & SSIM\\ 
    \midrule
    \textbf{BE-CALF}~\cite{liu2019calf} & NR & NR & 38.9271 & 0.9681 \\
    \textbf{BitNet}~\cite{byun2019bitnet} & 32.6832 & 0.9172 & 38.4822 & 0.9659 \\

    \midrule
    \textbf{BitMore D4}~\cite{punnappurath2021little} & 33.5089 & \textbf{0.9319} & 39.4171 & 0.9709 \\
    \textbf{Ours-4} & \textbf{33.7392}& 0.9315&\textbf{39.6566}&\textbf{0.9715}\\
    \midrule
    \textbf{BitMore D16}~\cite{punnappurath2021little} & 33.6679 & 0.9337 & 39.5185 & 0.9723 \\
    \textbf{Ours-16} & \textbf{33.8698}& \textbf{0.9331}&\textbf{39.6788}&\textbf{0.9725}\\
    \bottomrule
    \end{tabular}
    \label{tab:ko}
    }
\end{table}

\begin{table}[t!]
    \caption{Results on ESPL v2 dataset~\cite{kundu2015full}. NR denotes the score is `not reported' in the original paper.}
    \centering
    \small
    \scalebox{0.9}{
    \begin{tabular}{ccccc}
    \toprule
    \textbf{Method} & \multicolumn{2}{c}{\textbf{3-8 bit}} & \multicolumn{2}{c}{\textbf{4-8 bit}}\\
    & PSNR & SSIM & PSNR & SSIM\\ 
    \midrule
    \textbf{BE-CALF}~\cite{liu2019calf} & NR & NR & 38.4307 & 0.9479 \\
    \textbf{BitNet}~\cite{byun2019bitnet} & 32.5878 & 0.8717 & 38.2329 & 0.9399 \\
    \midrule
    \textbf{BitMore D4}~\cite{punnappurath2021little} & 33.1244 & \textbf{0.8981} & 39.3854 & \textbf{0.9532} \\
    \textbf{Ours-4} &\textbf{33.3526}&0.8976&\textbf{39.4163}&0.9474\\
    \midrule
    \textbf{BitMore D16}~\cite{punnappurath2021little} & 33.4685 & \textbf{0.9001} & \textbf{39.5312} & \textbf{0.9528} \\
    \textbf{Ours-16} & \textbf{33.5715}& 0.8950&39.4832&0.9455\\
    \bottomrule
    \end{tabular}
    \label{tab:es}
    }
\end{table}

\begin{table}[t!]
    \caption{Ablations of super-resolution modules across test datasets on 4-to-8 bit setting.}
    \label{tab:ds}
    \centering
    \small
    \resizebox{\linewidth}{!}{%
    \setlength{\tabcolsep}{0.7mm}{
    \begin{tabular}{@{}cccccccccc@{}}
    \toprule
    \multirow{2}{*}{\textbf{Baseline}} & \multirow{2}{*}{\textbf{SR}} & \multicolumn{2}{c}{\textbf{ESPL v2}} & \multicolumn{2}{c}{\textbf{TESTIMAGES}} & \multicolumn{2}{c}{\textbf{Sintel}} & \multicolumn{2}{c}{\textbf{Kodak}}          \\
                          &                     & PSNR              & SSIM             & PSNR           & SSIM                   & PSNR              & SSIM            & PSNR             & \multicolumn{1}{l}{SSIM} \\ \midrule
    \textbf{Ours-4}       & w/o                 & 39.1000           & \textbf{0.9531}           & 39.1970        & 0.9657                 & 39.7774           & 0.9734          & 39.0884          & 0.9691                   \\
    \textbf{Ours-4}       & EDSR       & \textbf{39.4163}  & 0.9474  & 39.7852        & \textbf{0.9697}        & \textbf{41.0086}  & \textbf{0.9793} & \textbf{39.6566} & \textbf{0.9715}          \\
    \textbf{Ours-4}       & RCAN                & 39.3637  & 0.9491  & \textbf{39.8309}        & 0.9686                 & 40.8340           & 0.9786          & 39.5500          & 0.9711                   \\ \bottomrule
    \end{tabular}%
    }}
\end{table}

\subsection{Qualitative Results}

\begin{figure}[htbp]
 \includegraphics[width=\linewidth]{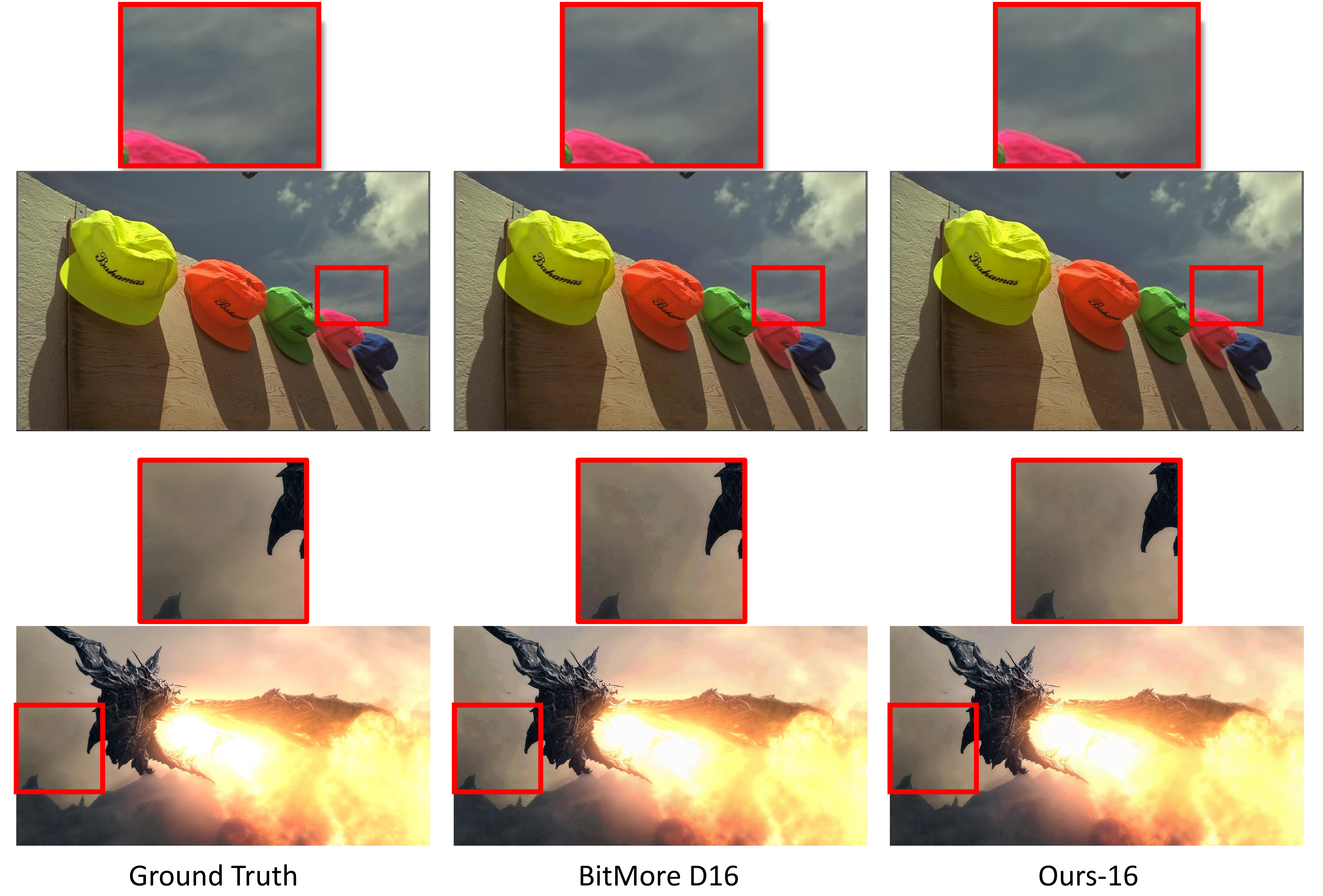}
    \caption{Visual comparison of our method versus BitMore. From left to right: Ground truth, Bitmore and Ours. Rows 2 and 4 present the full images, while rows 1 and 3 are close-ups.}
    \label{fig:combined}
    \vspace{-0.5cm}
\end{figure}

As illustrated in Fig. \ref{fig:combined}, the proposed model produces smoother images with significantly reduced banding artifacts. This demonstrates that incorporating a priori difference information obtained through super-resolution effectively aids in color bit-recovery, enabling enhanced attention to detail and improved color restoration.

\subsection{Ablation Study}

We evaluate the impact of super-resolution methods in the ablation study. Table~\ref{tab:ds} presents comparisons between configurations with and without SR modules in the feature extractor, as well as comparisons using different super-resolution methods in 4-to-8 bit settings. The ablation is conducted on our baseline with four IRA blocks (Ours-4). Without SR modules, only the inception-based block is used as the feature extractor.

On the ESPL v2 dataset, Ours-4 with EDSR achieves the highest PSNR (39.4163 dB), while Ours-4 with RCAN performs slightly better in SSIM (0.9491). For TESTIMAGES, Ours-4 with RCAN achieves the highest PSNR (39.8309 dB), while the EDSR version attains the best SSIM (0.9697). On the Sintel dataset, Ours-4 with EDSR demonstrates its superiority with a PSNR of 41.0086 dB and SSIM of 0.9793, outperforming our model with RCAN. These results highlight the effectiveness of the pretrained SR-based modules, and demonstrate that the generalization ability between priors extracted by models with different architectures.

\section{Conclusion}

In this paper, we propose enhancing the fidelity of the bit-depth recovery task using priors learned from super-resolution models. We design a super-resolution-based feature extractor combined with a lightweight bit-plane prediction network. Experimental results on four benchmark datasets demonstrate the superiority of our approach.

\newpage
\bibliographystyle{IEEEtran}
\bibliography{bib}

\begin{thebibliography}{10}
\providecommand{\url}[1]{#1}
\csname url@samestyle\endcsname
\providecommand{\newblock}{\relax}
\providecommand{\bibinfo}[2]{#2}
\providecommand{\BIBentrySTDinterwordspacing}{\spaceskip=0pt\relax}
\providecommand{\BIBentryALTinterwordstretchfactor}{4}
\providecommand{\BIBentryALTinterwordspacing}{\spaceskip=\fontdimen2\font plus
\BIBentryALTinterwordstretchfactor\fontdimen3\font minus \fontdimen4\font\relax}
\providecommand{\BIBforeignlanguage}[2]{{%
\expandafter\ifx\csname l@#1\endcsname\relax
\typeout{** WARNING: IEEEtran.bst: No hyphenation pattern has been}%
\typeout{** loaded for the language `#1'. Using the pattern for}%
\typeout{** the default language instead.}%
\else
\language=\csname l@#1\endcsname
\fi
#2}}
\providecommand{\BIBdecl}{\relax}
\BIBdecl

\bibitem{apple_support_sp770}
\BIBentryALTinterwordspacing
Apple, ``iphone 16 pro - technical specifications,'' 2024. [Online]. Available: \url{https://www.apple.com/iphone-16-pro/specs/}
\BIBentrySTDinterwordspacing

\bibitem{samsung_s10_specs}
\BIBentryALTinterwordspacing
Samsung, ``Galaxy z fold6 - technical specifications,'' 2024. [Online]. Available: \url{https://www.samsung.com/sg/smartphones/galaxy-z-fold6/specs/}
\BIBentrySTDinterwordspacing

\bibitem{karaimer2016software}
H.~C. Karaimer and M.~S. Brown, ``A software platform for manipulating the camera imaging pipeline,'' in \emph{European Conference on Computer Vision (ECCV)}, 2016, pp. 429--444.

\bibitem{ulichney1998pixel}
R.~A. Ulichney and S.~Cheung, ``Pixel bit-depth increase by bit replication,'' in \emph{Color Imaging: Device-Independent Color, Color Hardcopy, and Graphic Arts III}, vol. 3300, 1998, pp. 232--241.

\bibitem{cheng2009bit}
C.-H. Cheng, O.~C. Au, C.-H. Liu, and K.-Y. Yip, ``Bit-depth expansion by contour region reconstruction,'' in \emph{IEEE International Symposium on Circuits and Systems}, 2009, pp. 944--947.

\bibitem{mittal2012bit}
G.~Mittal, V.~Jakhetiya, S.~P. Jaiswal, O.~C. Au, A.~K. Tiwari, and D.~Wei, ``Bit-depth expansion using minimum risk based classification,'' in \emph{Visual Communications and Image Processing}, 2012, pp. 1--5.

\bibitem{wan20122d}
P.~Wan, O.~C. Au, K.~Tang, Y.~Guo, and L.~Fang, ``From 2d extrapolation to 1d interpolation: Content adaptive image bit-depth expansion,'' in \emph{IEEE International Conference on Multimedia and Expo}, 2012, pp. 170--175.

\bibitem{wan2016image}
P.~Wan, G.~Cheung, D.~Florencio, C.~Zhang, and O.~C. Au, ``Image bit-depth enhancement via maximum a posteriori estimation of ac signal,'' \emph{IEEE Transactions on Image Processing}, vol.~25, no.~6, pp. 2896--2909, 2016.

\bibitem{liu2018ipad}
J.~Liu, G.~Zhai, A.~Liu, X.~Yang, X.~Zhao, and C.~W. Chen, ``Ipad: Intensity potential for adaptive de-quantization,'' \emph{IEEE Transactions on Image Processing}, vol.~27, no.~10, pp. 4860--4872, 2018.

\bibitem{byun2019bitnet}
J.~Byun, K.~Shim, and C.~Kim, ``Bitnet: Learning-based bit-depth expansion,'' in \emph{Asian Conference on Computer Vision}, 2019, pp. 67--82.

\bibitem{su2019photo}
Y.~Su, W.~Sun, J.~Liu, G.~Zhai, and P.~Jing, ``Photo-realistic image bit-depth enhancement via residual transposed convolutional neural network,'' \emph{Neurocomputing}, vol. 347, pp. 200--211, 2019.

\bibitem{liu2019calf}
J.~Liu, W.~Sun, Y.~Su, P.~Jing, and X.~Yang, ``Be-calf: Bit-depth enhancement by concatenating all level features of dnn,'' \emph{IEEE Transactions on Image Processing}, vol.~28, no.~10, pp. 4926--4940, 2019.

\bibitem{punnappurath2021little}
A.~Punnappurath and M.~S. Brown, ``A little bit more: Bitplane-wise bit-depth recovery,'' \emph{IEEE Transactions on Pattern Analysis and Machine Intelligence}, vol.~44, no.~12, pp. 9718--9724, 2021.

\bibitem{zhao2019deep}
Y.~Zhao, R.~Wang, W.~Jia, W.~Zuo, X.~Liu, and W.~Gao, ``Deep reconstruction of least significant bits for bit-depth expansion,'' \emph{IEEE Transactions on Image Processing}, vol.~28, no.~6, pp. 2847--2859, 2019.

\bibitem{hou2017image}
X.~Hou and G.~Qiu, ``Image companding and inverse halftoning using deep convolutional neural networks,'' \emph{arXiv preprint arXiv:1707.00116}, 2017.

\bibitem{SR_survey}
Z.~Wang, J.~Chen, and S.~C.~H. Hoi, ``Deep learning for image super-resolution: A survey,'' \emph{IEEE Transactions on Pattern Analysis and Machine Intelligence}, vol.~43, no.~10, pp. 3365--3387, 2021.

\bibitem{Sintel}
\BIBentryALTinterwordspacing
X.~Foundation, ``Xiph.org,'' 2024. [Online]. Available: \url{http://www.xiph.org/}
\BIBentrySTDinterwordspacing

\bibitem{lim2017enhanced}
B.~Lim, S.~Son, H.~Kim, S.~Nah, and K.~Mu~Lee, ``Enhanced deep residual networks for single image super-resolution,'' in \emph{IEEE Conference on Computer vision and pattern recognition (CVPR) Workshops}, 2017, pp. 136--144.

\bibitem{conde2023perceptual}
M.~V. Conde, F.~Vasluianu, J.~Vazquez-Corral, and R.~Timofte, ``Perceptual image enhancement for smartphone real-time applications,'' in \emph{EEE/CVF Winter Conference on Applications of Computer Vision (WACV)}, 2023, pp. 1848--1858.

\bibitem{asuni2014testimages}
N.~Asuni, A.~Giachetti \emph{et~al.}, ``Testimages: a large-scale archive for testing visual devices and basic image processing algorithms.'' in \emph{Smart Tools \& Apps for Graphics (STAG)}, 2014, pp. 63--70.

\bibitem{Kodak}
\BIBentryALTinterwordspacing
E.~Kodak, ``Kodak lossless true color image suite,'' 1999. [Online]. Available: \url{http://r0k.us/graphics/kodak/}
\BIBentrySTDinterwordspacing

\bibitem{kundu2015full}
D.~Kundu and B.~L. Evans, ``Full-reference visual quality assessment for synthetic images: A subjective study,'' in \emph{IEEE International Conference on Image Processing (ICIP)}, 2015, pp. 2374--2378.

\bibitem{zhang2018image}
Y.~Zhang, K.~Li, K.~Li, L.~Wang, B.~Zhong, and Y.~Fu, ``Image super-resolution using very deep residual channel attention networks,'' in \emph{European Conference on Computer Vision (ECCV)}, 2018, pp. 286--301.

\bibitem{timofte2017ntire}
R.~Timofte, E.~Agustsson, L.~Van~Gool, M.-H. Yang, and L.~Zhang, ``Ntire 2017 challenge on single image super-resolution: Methods and results,'' in \emph{EEE Conference on Computer Vision and Pattern Recognition (CVPR) workshops}, 2017, pp. 114--125.

\end{thebibliography}


\end{document}